\documentclass[twocolumn,prl]{revtex4}
\usepackage{amsfonts}
\usepackage[T1]{fontenc}
\usepackage{amsmath,amsbsy,amssymb,graphicx}
\usepackage{amsmath}
\usepackage{amssymb}
\usepackage{graphicx}
\usepackage{times}

\let\mathbf=\boldsymbol

\begin{document}

\title{{\Large Compact Skyrmions, Merons and Bimerons in Thin Chiral
Magnetic Films}}
\author{Motohiko Ezawa}
\affiliation{Department of Applied Physics, University of Tokyo, Hongo 7-3-1, 113-8656,
Japan }

\begin{abstract}
A meron is a controversial topological excitation because it carries just one
half of the topological charge unit. It is believed that it is tightly
binded to another meron and cannot be observed by isolating it. We
present a counter example, investigating the 2-dimensional nonlinear sigma
model together with the Dzyaloshinskii-Moriya interaction, where topological
excitations are merons, bimerons and skyrmions. They behave as if they were
free particles since they are electrically neutral. A prominent feature
is that the topological charge density is strictly confined within compact
domains. We propose an analytic approach for these compact excitations, and
construct a phase diagram. It is comprised of the helix, meron,
skyrmion-crystal, skyrmion-gas and ferromagnet phases. It explains quite
well the experimental data recently performed in chiral magnets such as MnSi
and FeCoSi thin films, as verifies that merons are surely basic topological
excitations in the system.
\end{abstract}

\date{\today }
\maketitle

Topological excitations are endlessly fascinating. They are constantly under
investigations in all branches of physics. Well known examples are vortices
and skyrmions\cite{Skyrmion}. A fantastic topological object is a meron: It
cannot exist by itself since it carries only one half of the topological
charge unit. A meron was originally invented as a half-instanton in the
context of quark confinement in particle physics\cite{Callan77}. Later it
was introduced as a half-skyrmion in order to account for a certain
anomalous behavior in bilayer quantum Hall effects\cite{Moon95B}. Though the
meron is a theoretically useful concept, it remains to be a long-standing
problem whether it exists in reality. There is no clear evidence for its
existence in spite of experimental endeavors\cite{Skyrmion}.

Magnetic thin films are ideal systems to investigate and test various
intriguing ideas on topological excitations. Indeed, a skyrmion crystal\cite%
{Mohlbauer} as well as a single skyrmion\cite{Yu} have been observed in
chiral magnets such as MnSi and FeCoSi thin films. Furthermore, magnetic
domains observed in ferromagnets such as a TbFeCo thin film\cite{Ogasawara}
are shown to be giant skyrmions\cite{EzawaGiant} as large as $\sim 1\mu m$.
In this paper we point out that merons are also detectable as almost
isolated objects.

\begin{figure}[t]
\centerline{\includegraphics[width=0.45\textwidth]{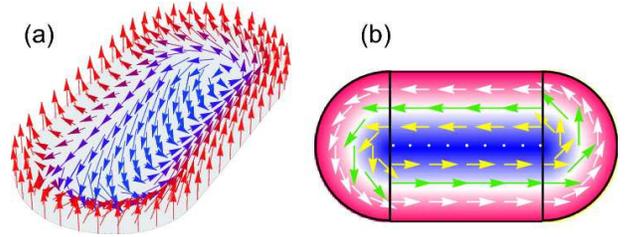}}
\caption{{}(Color online) (a) The spin texture of a compact bimeron in a
thin chiral magnetic film. (b) A compact bimeron is composed of two
half-disk domains and a rectangular stripe domain. Spins are pointed up on
the boundary and down deep inside of the bimeron. They are pointed forward
and backward in the recutangular part, and twisting circularly in the
half-disk part. The topological charge density is nonvanishing only in the
half-disk parts, each of which has $Q_{\text{sky}}=1/2$. The half-disk part
is identified as a meron. The spin texture of a compact skyrmion is obtained
simply by removing the recutangular part and by patching the two half-disk
domains. }
\label{FigBimeron}
\end{figure}

The ground state of a chiral magnet is a helical state in the absence of
external magnetic field. The spin texture of the helical state has a
stripe-domain structure, where the width of a stripe has a fixed value
determined by sample parameters. A stripe breaks into pieces as the magnetic
field increases. By calculating the topological charge density, we show that
the endpoint of a broken stripe has the topological charge $Q_{\text{sky}%
}=1/2$. It is natural to identify them as merons. Finite-length stripes are
bimerons [Fig.\ref{FigBimeron}], among which the shortest ones are skyrmions.

When we talk about skyrmions in the 2-dimensional space, it is implicit to
assume a Belavin-Polyakov skyrmion. Its spin texture approaches the
ground-state value only polynomially at large distance. On the contrary, a
skyrmion must be strictly compact in the chiral magnet since it is embedded
within a stripe-domain structure. We propose an analytic scheme to explore
compact skyrmions, merons and bimerons. Being electrically neutral
excitations, they behave as if they were free particles. Based on this
observation we construct a phase diagram of the chiral ferromagnet. It
explains quite well the recent experimental data carried out in a FeCoSi
thin film\cite{Yu}. This fact shows that the identification of merons is
justified not only theoretically but also experimentally.

Our system is the 2-dimensional plane described by the nonlinear O(3) sigma
model $H_{J}$ with easy axis anisotropy,%
\begin{equation}
H_{J}=\frac{1}{2}\Gamma \int \!d^{2}x\,[\left( \partial _{k}\mathbf{n}%
\right) \left( \partial _{k}\mathbf{n}\right) -\xi ^{-2}\left( n_{z}\right)
^{2}],  \label{SigmaModel}
\end{equation}%
and the Dzyaloshinskii-Moriya interaction (DMI),%
\begin{equation}
H_{\text{DM}}=D\int \!d^{2}x\,\mathbf{n}(\mathbf{x})\cdot \left( \nabla
\times \mathbf{n}(\mathbf{x})\right) ,  \label{DMI}
\end{equation}%
where $\Gamma =(1/2)zS^{2}J$ is the exchange energy ($z$ denotes the number
of nearest neighbors, $S$ the spin per atom, $J$ the exchange constant), $%
\xi $ is the single-ion easy-axis-anisotropy constant, and $\mathbf{n}%
=(n_{x},n_{y},n_{z})$ is a classical spin field of unit length. The DMI term
breaks the chiral symmetry explicitly. We introduce the magnetic field $h$
perpendicular to the plane with the Zeeman energy $\Delta _{Z}=Sg\mu _{B}\mu
_{0}h$, 
\begin{equation}
H_{Z}=-(\Delta _{Z}/a^{2})\int d^{2}x\,n_{z}(\mathbf{x}),  \label{ZeemaInter}
\end{equation}%
where $a$ is the lattice constant.

\begin{figure}[t]
\centerline{\includegraphics[width=0.48\textwidth]{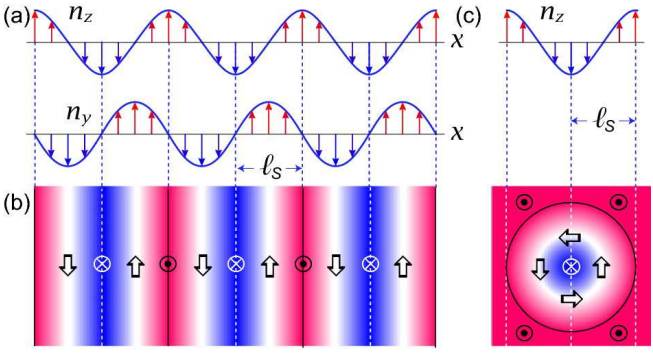}}
\caption{(Color online) (a) Illustration of the helical state solution $%
n_{i}(x)$ described by the Jacobian elliptic function (\protect\ref%
{SolitLatti}). Here we have taken $\protect\kappa =0.3$. (b) Illustration of
the spin structure with an alternating up-down and forward-backward
stripe-domains. Spins are strictly pointed up or down on the vertical solid
or dotted lines. As one stripe we consider the region sandwiched by two
solid lines. (c) Illustration of a compact skyrmion described by the
Jacobian elliptic function (\protect\ref{SkyrmField}) . It is constructed in
such a way that any cross section passing through the center agrees with the
cross section of the stripe. The topological charge is strictly confined
within the solid circle.}
\label{FigSL}
\end{figure}

We have emphasized previously\cite{EzawaGiant} the importance of the
magnetic dipole-dipole interaction (DDI). However, since the DDI constant is
very small compared to the DMI constant, we may ignore it in determining the
magnetic structure of a chiral magnet. Indeed, the typical size determined
by the DMI is of the order of $40$nm, where the DDI is negligible.

We start with a study of the system in the absence of external magnetic
field. The ferromagnetic spin state $\mathbf{n}=(0,0,\pm 1)$ is a solution
of the Hamiltonian $H_{J\text{-DM}}=H_{J}+H_{\text{DM}}$ with the energy%
\begin{equation}
E_{\text{homo}}=-\Gamma /2\xi ^{2}.  \label{EnergHomo}
\end{equation}%
However, in general, this is not the ground state. It is easy to prove that
the Hamiltonian $H_{J\text{-DM}}$ allows one-dimensional periodic solutions,
among which the one that minimizes the DMI term (\ref{DMI}) is given by%
\begin{align}
n_{x}(x)=& 0,\quad n_{y}(x)=\text{cn}\left( \frac{x-\ell _{S}/2}{\kappa \xi }%
,\kappa ^{2}\right) ,  \notag \\
n_{z}(x)=& \text{sn}\left( \frac{x-\ell _{S}/2}{\kappa \xi },\kappa
^{2}\right) ,  \label{SolitLatti}
\end{align}%
in terms of the Jacobian elliptic functions [Fig.\ref{FigSL}(a)], where $%
\kappa $ is an integration constant with $0\leq \kappa ^{2}\leq 1$. The
periodicity of $\sigma (x)$ is $2\ell _{S}$ with%
\begin{equation}
2\ell _{S}\equiv 4\kappa \xi K(\kappa ^{2}),  \label{PerioSL}
\end{equation}%
where $K(\kappa ^{2})$ is the complete elliptic integral of the first kind.
The periodic state (\ref{SolitLatti}) has an alternating up-down and
forward-backward spin-stripe structure, as illustrated in Fig.\ref{FigSL}%
(b). It may be called the anisotropic helical state. Note that $n_{z}(-\ell
_{S})=1$, $n_{z}(0)=-1$, $n_{z}(\ell _{S})=1$. As one stripe we consider the
region whose width is $2\ell _{S}$ with the center line being given by $%
n_{z}(x)=-1$.

By substituting (\ref{SolitLatti}) into the Hamiltonian $H_{J\text{-DM}}$,
the energy of the helical state is analytically calculable,%
\begin{equation}
E_{\text{helix}}=L_{x}L_{y}\left[ \frac{\Gamma }{2\xi ^{2}}\left( \frac{2}{%
\kappa ^{2}}\frac{E(\kappa ^{2})}{K(\kappa ^{2})}-\frac{1}{\kappa ^{2}}%
+1\right) -\frac{\pi D}{2\kappa \xi K(\kappa ^{2})}\right] ,
\end{equation}%
where $L_{x}$ and $L_{y}$ are the sample size in the $x$ and $y$ direction.
We determine the parameter $\kappa $ by minimizing $E_{\text{helix}}$ with
respect to $\kappa $. It is given by solving $2E(\kappa ^{2})=\kappa \pi
D\xi /\Gamma $. Provided $D\xi \gg \Gamma $, it is solved as%
\begin{equation}
\kappa =\frac{\Gamma }{D}\frac{1}{\xi }-\frac{1}{4}\left( \frac{\Gamma }{D}%
\frac{1}{\xi }\right) ^{3}+\cdots ,  \label{EqA}
\end{equation}%
and the energy of the helical state is%
\begin{equation}
E_{\text{helix}}=L_{x}L_{y}\left[ -\frac{D^{2}}{2\Gamma }+\frac{\Gamma }{%
2\xi ^{2}}+\cdots \right] .
\end{equation}%
We compare this with that of the homogeneous state (\ref{EnergHomo}). We
find that the helical state has a lower energy than that of the homogeneous
state if $D\xi >\sqrt{2}\Gamma $. It is interesting that the helical state
does not realize in the sample when the anisotropy is too large.

\begin{figure}[t]
\centerline{\includegraphics[width=0.33\textwidth]{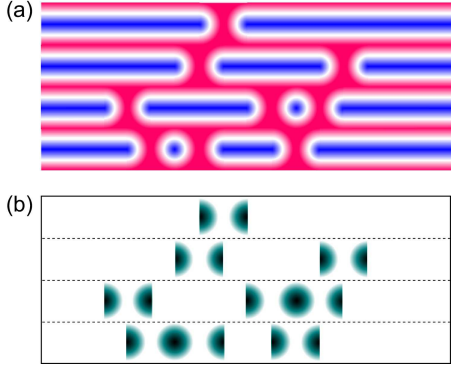}}
\caption{{}(Color online) (a) Illustration of compact skyrmions, merons and
bimerons embedded in the stripe-domain structure. (b) Illustration of the
topological charge density confined within compact domains. Dotted lines
show the boundaries of stripes.}
\label{FigMeronPhase}
\end{figure}

We switch on the external magnetic field. The Zeeman effect enforces the
increase of the up-spin region. However, it is impossible to increase only
the width of the up-spin part of the stripe, which is fixed to be $\ell _{S}$%
. The simple way is to split a stripe into two stripes, since it increases
up-spin region [Fig.\ref{FigMeronPhase}(a)]. Let us cut one stripe at $y=0$
and then put a cap so that the spin field on the cross section smoothly
approaches the up-spin value at the boundary in order to optimize the energy
[Fig.\ref{FigBimeron}(b)]. The spin texture of the cap must be given by%
\begin{align}
n_{x}(r,\theta )& =-\text{cn}\left( \frac{r-\ell _{S}/2}{\kappa \xi },\kappa
^{2}\right) \sin \theta ,\quad  \notag \\
n_{y}(r,\theta )& =\text{cn}\left( \frac{r-\ell _{S}/2}{\kappa \xi },\kappa
^{2}\right) \cos \theta ,  \notag \\
n_{z}(r,\theta )& =\text{sn}\left( \frac{r-\ell _{S}/2}{\kappa \xi },\kappa
^{2}\right) ,  \label{SkyrmField}
\end{align}%
for the half-disk region ($r\leq \ell _{S},0\leq \theta \leq \pi $) in the
cylindrical coordinate, since it agrees with (\ref{SolitLatti}) at $y=0$,
where $\cos \theta =1$.

A stripe may be broken into three stripes with one finite-length stripe [Fig.%
\ref{FigMeronPhase}(a)]. The spin structure of a finite-length stripe is
illustrated in Fig.\ref{FigBimeron}. The shortest stripe is a cylindrical
symmetric domain [Figs.\ref{FigMeronPhase}(a) and \ref{FigSL}(c)], whose
spin texture is described by (\ref{SkyrmField}) for the disk region ($r\leq
\ell _{S},0\leq \theta \leq 2\pi $).

The use of a continuum approximation and of classical fields to represent
the spins is justified as far as we analyze phenomena whose characteristic
wavelength is much larger than the lattice constant. In this regime there
exists the topologically conserved charge, that is the Pontryagin number, $%
Q_{\text{sky}}=\int \!d^{2}x\,\rho _{\text{sky}}(\mathbf{x})$, with the
topological charge density%
\begin{equation}
\rho _{\text{sky}}(\mathbf{x})=-{\frac{1}{8\pi }}\sum_{ij}\varepsilon _{ij}%
\mathbf{n}(\mathbf{x})\cdot \left( \partial _{i}\mathbf{n}(\mathbf{x})\times
\partial _{j}\mathbf{n}(\mathbf{x})\right) ,  \label{PontrNumbe}
\end{equation}%
where $i,j$ run over $x,y$ with $\varepsilon _{ij}$ being the completely
antisymmetric tensor. We are able to determine the topological charge
density for various spin textures.

First, it is trivial to see that the stripe configuration (\ref{SolitLatti})
has no topological density. Then, calculating it for the half-disk
configuration (\ref{SkyrmField}) with $0\leq \theta \leq \pi $, we find that 
$Q_{\text{sky}}=1/2$. Similarly we find $Q_{\text{sky}}=1$ for the
cylindrical symmetric configuration (\ref{SkyrmField}) with $0\leq \theta
\leq 2\pi $. Clearly we can identify them as a meron and a skyrmion,
respectively. A prominent feature is that the topological charge is strictly
confined within a compact domain. Hence we may call them a compact skyrmion
and so on. In general, there appear a variety of topological excitations
[Fig.\ref{FigMeronPhase}(a)]. We have illustrated the corresponding
topological charge density in Fig.\ref{FigMeronPhase}(b). Let us refer to
this regime of topological excitations as the meron phase, since the basic
excitation is a meron that appears at the endpoint of a stripe.

We now estimate the energy of a topological configuration. The Jacobian
elliptic functions are well approximated by the sinusoidal functions when $%
\kappa $ is not close to $1$, say, $\kappa \lesssim 0.5$. Namely, we may
approximated (\ref{SolitLatti}) by $n_{x}(x)=0$, $n_{y}(x)=\sin (kx)$ and $%
n_{z}(x)=-\cos (kx)$, where $k=1/\kappa \xi $. This is the well-known
expression for the helical ground state in the isotropic system, which is
the limit $\xi \rightarrow \infty $ with $k$ being fixed. The spin texture
of the cap (\ref{SkyrmField}) is approximated by%
\begin{align}
n_{x}(r,\theta )& =-\sin (kr)\sin \theta ,\quad n_{y}(r,\theta )=\sin
(kr)\cos \theta ,  \notag \\
n_{z}(r,\theta )& =-\cos (kr),  \label{StripField}
\end{align}%
for $r\leq \pi /2k$. We consider the isotropic system for simplicity.

With the use of the meron configuration (\ref{StripField}), by integrating
the total Hamiltonian $H=H_{J}+H_{\text{DM}}+H_{Z}$, it is straightforward
to calculate the energy gain when a stripe is broken into two stripes,%
\begin{equation}
\Delta E_{\text{merons}}=\ell _{S}^{2}\left[ \left( \frac{4}{\sqrt{3}}-\frac{%
\pi }{2}\right) \frac{D^{2}}{\Gamma }-\left( \frac{4}{\pi }+\frac{8}{\sqrt{3}%
}-\pi \right) \Delta _{Z}\right] .  \label{EnergMerons}
\end{equation}%
We may call it the creation energy of a meron pair. The energy gain when a
skyrmion emerges in a stripe is just twice of $\Delta E_{\text{merons}}$. It
follows from (\ref{EnergMerons}) that $\Delta E_{\text{merons}}<0$ if $%
\Delta _{Z}>\Delta _{Z}^{\text{Helix-SkX}}$ with%
\begin{equation}
\Delta _{Z}^{\text{Helix-SkX}}=0.27D^{2}/\Gamma .
\end{equation}%
Since the energy gain is negative, all stripes are spontaneously broken into
a maximum number of skyrmions for $\Delta _{Z}>\Delta _{Z}^{\text{Helix-SkX}%
} $, which would lead to the formation of a skyrmion crystal (SkX). Namely, $%
\Delta _{Z}^{\text{Helix-SkX}}$ is the phase-transition point between the
helix and SkX phases. It is concluded that there exists no meron phase at
zero temperature.

The SkX has been discussed in literature\cite{Bog,HanNagaoka}, though
Belavin-Polyakov skyrmions are assumed on all lattice points with a certain
cutoff. Note that a skyrmion with a cutoff does not have the correct
topological charge. We can follow their arguments to study the SkX with the
use of compact skyrmions without any problem.

On the other hand, the ferromagnet (FM) phase appears in sufficiently strong
external magnetic field, which has only the Zeeman energy, $E_{\text{FM}%
}=-\left( L_{x}L_{y}/a^{2}\right) \Delta _{Z}$. By comparing this with (\ref%
{EnergMerons}), it follows that the critical Zeeman energy is%
\begin{equation}
\Delta _{Z}^{\text{SkX-FM}}=0.84D^{2}/\Gamma ,  \label{CritiZeema}
\end{equation}%
so that the FM phase appears for $\Delta _{Z}>\Delta _{Z}^{\text{SkX-FM}}$.

We proceed to construct the phase diagram in the plane of temperature and
magnetic field. We have it already at zero temperature, where there exists
only the helix, SkX and FM phases. The meron phase appears at finite
temperature, since its entropy is much larger than that of the helix or SkX
phase.

\begin{figure}[t]
\centerline{\includegraphics[width=0.49\textwidth]{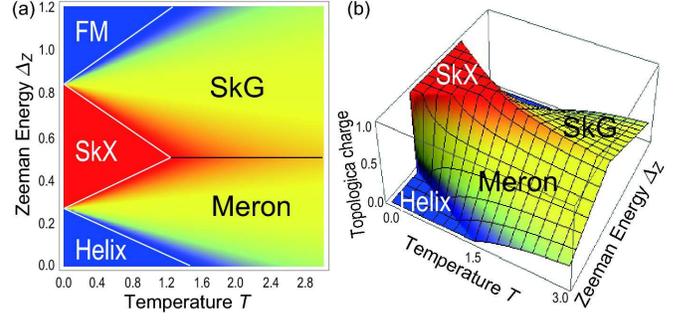}}
\caption{{}(Color online) (a) Phase diagram. The horizontal axis is the
temperature $T$ in unit of $D^{2}/k_{\text{B}}\Gamma $, while the vertical
axis is the Zeeman energy $\Delta _{Z}$ in unit \ of $D^{2}/\Gamma $.\ (b)
Average topological charge density in various phases. }
\label{FigPhasePT}
\end{figure}

In determining the boundary between the helix and meron phases, the basic
excitation is a pair of merons in the helix phase [Fig.\ref{FigMeronPhase}%
(b)]. It appears when a stripe is broken into two pieces with the excitation
energy $\Delta E_{\text{merons}}$ being given by (\ref{EnergMerons}). Let $N$
be the maximum number of the topological charges (the maximum number of
compact skyrmions) that the system can accommodate. When the topological
charge of the system is $n$, the entropy is given by $S=\ln N!/n!\left(
N-n\right) !$. Then, the free energy at temperature $T$ is given by $F=E_{%
\text{helix}}+n\Delta E_{\text{merons}}-TS$. Letting $N\rightarrow \infty $,
we obtain the formula for the free energy density,%
\begin{equation}
f=\varepsilon _{\text{helix}}+q\Delta E_{\text{merons}}+\frac{k_{\text{B}}T}{%
2}\left( q-\frac{1}{2}\right) ^{2},  \label{FreeEnergDensi}
\end{equation}%
where $f=F/N$, $\varepsilon _{\text{helix}}=E_{\text{helix}}/N$, and $q=n/N$
is the average topological charge density ($0\leq q\leq 1$). It is easy to
minimize the free energy density $f$ with respect to $q$. Since $f$ is
quadratic in $q$, it yields two lines starting from the helix-SkX
phase-transition point at $T=0$. They determines the boundary between the
helix-meron phases and the boundary between the meron-SkX phases, as
illustrated in Fig.\ref{FigPhasePT}.

The SkX melts into a skyrmion gas (SkG) at higher temperature. We can make a
similar argument to derive the boundary between the SkX and SkG phases,
where the basic object is a skyrmion in the SkX phase with the excitation
energy being 
\begin{equation}
\Delta E_{\text{sky}}=-\ell _{S}\left( \frac{\pi D^{2}}{2\Gamma }+\left( 
\frac{4}{\pi }-\pi \right) \Delta _{Z}\right) .
\end{equation}
We reach at the same formula as (\ref{FreeEnergDensi}) with the replacement
of $\varepsilon _{\text{helix}}$ by $\varepsilon _{\text{SkX}}$, and $\Delta
E_{\text{merons}}$ by $\Delta E_{\text{sky}}$. In this way we obtain the
boundary between the SkX-SkG phases and the boundary between the SkG-FM
phases, as illustrated in Fig.\ref{FigPhasePT}. Finally, the boundary
between the meron and SkG phases is given by $\Delta _{Z}^{\text{meron-SkG}%
}=D^{2}/2\Gamma $ for $T>4D^{2}/\pi k_{\text{B}}\Gamma $ by comparing their
free energies.

The phase diagram thus constructed is characterized by the topological
charge density $q$ as follows:%
\begin{equation}
q=\left\{ 
\begin{array}{llll}
0 & \text{for} & \Delta E_{i}>\frac{1}{2}k_{\text{B}}T & \text{(Helix, FM)}
\\ 
\frac{1}{2}-\frac{\Delta E_{i}}{T} & \text{for} & \left\vert \Delta
E_{i}\right\vert <\frac{1}{2}k_{\text{B}}T & \text{(Meron, SkG)} \\ 
1 & \text{for} & -\Delta E_{i}>\frac{1}{2}k_{\text{B}}T & \text{(SkX)}%
\end{array}%
\right. ,
\end{equation}%
where $\Delta E_{i}$ stands for $\Delta E_{\text{merons}}$ or $\Delta E_{%
\text{sky}}$. This phase diagram [Fig.\ref{FigPhasePT}] captures the
essential feature of those obtained experimentally and by a Monte Carlo
simulation\cite{Yu}, although they have made no distinction between the
meron and SkG phases. We note that the topological charge density $q$ is
observable by measuring the Hall conductance $\sigma _{xy}$ of the
topological current\cite{THE,THEx}, $\sigma _{xy}\varpropto q$.

We remark that all topological excitations have topological charges of the
same sign in magnetic thin films. Thus, apparently, a stripe with $Q_{\text{%
sky}}=0$ breaks into a number of bimerons each of which carries $Q_{\text{sky%
}}=1$. One may wonder how the topological conservation holds in these
systems. We have pointed out previously\cite{EzawaGiant} that a single
skyrmion can be created in a thin ferromagnet film by destroying the
magnetic order within a tiny spot with the use of photoirradiation. Indeed,
the topological charge is well defined only when we can describe the spin
system by continuous classical fields. It means that we can break the
topological conservation by controlling the system at the lattice-constant
scale or at sufficiently high temperature. Thus, in order to create
topological excitations, it is necessary to cool the sample from high
temperature. Once they are created, their stability is guaranteed
topologically. This would be the basic reason why a rich variety of stable
spin textures have been observed\cite{Yu}.

We have proposed a new concept of compact topological excitations together
with their analytic expressions, based on which we have explored thin chiral
magnetic films. Having identified merons as endpoints of stripes, we have
pointed out that they have already been observed almost as isolated objects
experimentally. The reason why merons can be observed is that they bear no
electric charge in magnetic films. Hence, the length of a stripe, which is
nothing but a bimeron, is a zero-energy mode. This allows a meron to behave
effectively as an isolated topological object though it carries only one
half of the topological charge unit. It is worthwhile to search for such
merons in other branches of physics.

I am deeply indebted to Y. Tokura, Y. Onose, X.Z. Yu and A. Rosch for
illuminating discussions and for informing me as to experimental details. I
am very much grateful to N. Nagaosa and J.H. Han for fruitful discussions on
the subject. This work was supported in part by Grants-in-Aid for Scientific
Research from the Ministry of Education, Science, Sports and Culture No.
22740196 and 21244053.

\end{document}